\documentclass[
a4paper,
aps,
prd,
preprint,
amsmath,
amssymb,
showpacs,
nofootinbib]{revtex4-1}
\usepackage[pdftex]{graphicx}
\usepackage{bm}
\usepackage{hyperref}
\usepackage{url}
\usepackage{comment}

\begin{document}

\title{Effects of initial spatial phase in radiative
       neutrino pair emission}

\date{\today}

\author{Minoru~Tanaka}
\email{tanaka@phys.sci.osaka-u.ac.jp}
\affiliation{Department of Physics, Graduate School of Science, 
             Osaka University, Toyonaka, Osaka 560-0043, Japan}

\author{Koji~Tsumura}
\affiliation{Department of Physics, Kyoto University, Kyoto 606-8502, Japan}

\author{Noboru~Sasao}
\author{Satoshi~Uetake}
\author{Motohiko~Yoshimura}
\affiliation{Research Institute for Interdisciplinary Science, 
             Okayama University, Tsushima-naka 3-1-1, Kita-ku,
             Okayama 700-8530, Japan}

\begin{abstract}
We study radiative neutrino pair emission in deexcitation process of atoms
taking into account coherence effect in a macroscopic target system.
In the course of preparing the coherent initial state to enhance the rate,
a spatial phase factor is imprinted in the macroscopic target.
It is shown that this initial spatial phase changes the kinematics
of the radiative neutrino pair emission. We investigate effects of the initial
spatial phase in the photon spectrum of the process.
It turns out that the initial spatial phase provides us significant
improvements in exploring neutrino physics such as the Dirac-Majorana
distinction and the cosmic neutrino background.
\end{abstract}


\preprint{KUNS-2703, OU-HET-950}

\maketitle

\section{Introduction}
\begin{figure}
 \centering
 \includegraphics[width=0.4\textwidth]{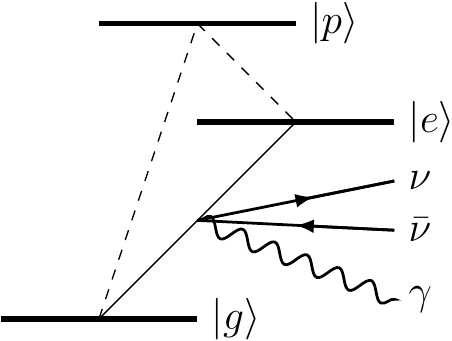}
 \caption{A schematic description of RENP. 
          The intermediate state is denoted by $|p\rangle$.}
 \label{Fig:RENPscheme}
\end{figure}

Radiative emission of neutrino pair (RENP) from atoms or molecules 
has been considered as a novel tool in neutrino physics~\cite{Fukumi2012a}.
The standard model of particle physics predicts that an excited state 
$|e\rangle$ of an atom deexcites to a ground (or lower-energy) state 
$|g\rangle$ by emitting a neutrino-antineutrino pair and a photon, 
$|e\rangle\to|g\rangle+\gamma+\nu\bar\nu$, as depicted in 
Fig.~\ref{Fig:RENPscheme}.
A rate enhancement mechanism using coherence in a macroscopic ensemble
of atoms is proposed in order to overcome the rate suppression
\cite{Yoshimura:2008a}.
This macrocoherent enhancement mechanism is experimentally confirmed
in the QED process in which the neutrino pair is replaced by
another photon, $|e\rangle\to|g\rangle+\gamma+\gamma$ 
(paired superradiance, PSR~\cite{YoshimuraSasaoTanaka2012a}), and
a rate amplification of $O(10^{18})$ is achieved using parahydrogen
molecules~\cite{Miyamoto2014a,Miyamoto2015a}.

In atomic deexcitation processes, the energy of the system is conserved
but the momentum is not as far as the recoil of the atom is
neglected. A peculiar nature of the macrocoherent enhancement is
that the kinematic configurations in which the momenta
of outgoing particles are balanced are selectively amplified.
In the RENP above, assigning momenta as 
$|e\rangle\to|g\rangle+\gamma(p_\gamma)+\nu_j(p)\bar\nu_i(p')$,
one may write the total amplitude as
\begin{equation}\label{Eq:Amp}
\text{Amp.}\propto\sum_a e^{-i(\bm{p}_\gamma+\bm{p}+\bm{p'})\cdot\bm{x}_a}
\simeq\frac{N}{V}(2\pi)^3\delta^3(\bm{p}_\gamma+\bm{p}+\bm{p'})\,,
\end{equation}
where $a$ and $\bm{x}_a$ denote an atom and its position respectively, 
the summation runs over $N$ atoms in the macroscopic target of volume $V$,
and the exponential factor represents the plane waves of the emitted particles.
All relevant atoms are supposed in an identical state
including their phases, so that the amplitudes of atoms are coherently
summed in Eq.~\eqref{Eq:Amp}. (See below for details.)
Thus, the macrocoherence implies momentum conservation as well as 
the energy conservation in atomic processes. 
The four-momentum conservation is represented as
$P^\mu=p_\gamma^\mu+p^\mu+p'^\mu$ with $(P^\mu):=(E_{eg},\bm{0})$,
where $E_{eg}:=E_e-E_g$ is the energy difference between the two atomic states. 

The four-momentum $P^\mu$ is regarded as that of ``a parent particle''
at rest and then the kinematics of the macrocoherently amplified RENP
is equivalent to that of three-body decay of this virtual parent particle.
Thus the photon spectrum in the RENP enhanced by the macrocoherence
is expected to be sensitive to neutrino masses.
Extending the standard model to include neutrino masses and mixings, 
some RENP spectra are calculated
\cite{Fukumi2012a,DinhPetcovSasaoTanakaYoshimura2012a,YoshimuraSasao2013a}.
It is shown that the RENP spectrum gives the information on
unknown neutrino properties, such as absolute masses,
Dirac/Majorana nature and Majorana phases
(if the neutrinos are Majorana fermions).
Quantitatively, the fine sensitivity to neutrino properties related to
masses is owing to the fact that the invariant mass of the parent particle,
$\sqrt{P^2}=E_{eg}$, is typically $O(1)$ eV and closer to the neutrino 
mass scale $\sim O(0.1)$ eV than other neutrino experiments. 

In the present work, we further pursue this kinematic advantage of RENP
in order to increase sensitivity to neutrino properties taking 
the effect of initial spatial phase (ISP) into consideration.
As described in the following, the ISP imprinted in a macroscopic target
works as a spatial component of the momentum $P^\mu$ of the virtual parent
particle, so that the invariant mass becomes smaller than $E_{eg}$.
We call the RENP process from a macroscopic target with the ISP as 
boosted RENP.

In Sec.~\ref{Sec:Kinematics}, we describe the spatial phase given to
a macroscopic target in the preparation process of initial coherent state
by two-photon absorption. 
The kinematics of the boosted RENP is also examined. 
We present a rate formula of the boosted RENP in Sec.~\ref{Sec:Rate}.
Section \ref{Sec:DM} is devoted to our numerical results on 
enhanced power of Dirac-Majorana distinction in the boosted RENP
as well as increased sensitivity to Majorana phases. 
In Sec.~\ref{Sec:CNB}, we discuss possible improvement in detecting
the cosmic neutrino background with spectral distortion in RENP.
Our conclusion is given in Sec.~\ref{Sec:Conclusion}.

\section{\label{Sec:Kinematics} Initial spatial phase and kinematics}
Prior to describing the ISP and its implication, we recapitulate the nature
of initial atomic states required for RENP.
In Eq.~\eqref{Eq:Amp}, the amplitude of each atom is assumed to
interfere with each other in addition to the phase matching by
the momentum conservation. The interference among atoms is
realized if the initial state of each atom is a superposition of
$|e\rangle$ and $|g\rangle$, e.g. 
$|\psi\rangle:=(|e\rangle+|g\rangle)/\sqrt{2}$. 

Suppose that $N$ atoms are in the initial state 
$\Pi_{a=1}^N|\psi\rangle_a$.
A deexcitation process such as the RENP is schematically expressed by
the lowering operator $\sum_{a=1}^N|g\rangle_a\,{_a\!\langle e|}$. 
The action of this operator on the above initial state gives 
the wave function of the final state, 
$(1/\sqrt{2})\sum_{i=1}^N|\psi\rangle_1\cdots|g\rangle_i\cdots|\psi\rangle_N$.
One finds that all the states in the summation interfere with
each other. Therefore the deexcitation rate, which is proportional to 
the square of the final wave function, behaves as $N^2$ when $N$ is
large.

Atomic states including the one like $|\psi\rangle$ are conveniently 
described by the density operator $\hat\rho$. 
The offdiagonal element, $\langle e|\hat\rho|g\rangle$, provides
the coherence that leads to the above $N^2$ behavior.
An initial atomic state with such coherence in a target can be prepared 
by the two-photon absorption process, $\gamma_1+\gamma_2+|g\rangle\to|e\rangle$,
with high quality lasers.
We note that
an electric dipole forbidden metastable state $|e\rangle$ is
preferable as an initial state in order to suppress ordinary fast QED 
deexcitation processes. Thus the single photon excitation is disfavored
as well.
In the numerical illustration in Sec.~\ref{Sec:DM}, 
we consider a $0^-\to 0^+$ transition of ytterbium, 
in which all multipole processes of single photon are forbidden.
\footnote{
The E1$\times$E1 two-photon process is prohibited by the parity.
The most serious QED process competing with
the RENP is the macrocoherently amplified three-photon emission.
It is shown that this three-photon process can be controlled with
a metal or photonic crystal 
waveguide~\cite{YoshimuraSasaoTanaka2015a,TanakaTsumuraSasaoYosimura2017a}.}

In the two-photon absorption process, the energy is conserved 
as $\omega_1+\omega_2=E_{eg}$, where $\omega_{1(2)}$ is the energy of 
$\gamma_{1(2)}$, but the momentum need not. 
Instead, the sum of the photon momenta, 
$\bm{p}_{eg}:=\bm{k}_1+\bm{k}_2$, where $\bm{k}_{1(2)}$ represents 
the momentum of $\gamma_{1(2)}$, is memorized in the resulting state
of the macroscopic target as a spatial phase factor. Therefore,
in the continuum approximation, which is valid for high density targets,
one may write $\langle e|\hat\rho|g\rangle$ of the prepared target state
as a product of the slowly varying function of the position $\bm{x}$ 
and the ISP factor of rapid oscillation,
\begin{equation}\label{Eq:Coherence}
\langle e|\hat\rho|g\rangle
=n\rho_{eg}(\bm{x})e^{i\bm{p}_{eg}\cdot\bm{x}}\,,
\end{equation}
where $n$ is the number density of target\footnote{The target number
density may be a function of the position. Here we assume a uniform
target for simplicity.}
and $\rho_{eg}(\bm{x})$ represents the envelope.
This is called as the slowly varying envelope approximation
in the literature. 
We note that, in Eq.~\eqref{Eq:Amp}, the atomic state is implicitly 
assumed to be prepared with the parent four-momentum
$(E_{eg},\bm{0})$ in the scheme of counter-propagating irradiation
of identical two lasers, $\omega_1=\omega_2=E_{eg}/2$ and 
$\bm{p}_{eg}=\bm{k}_1+\bm{k}_2=0$.

It is apparent that $\bm{p}_{eg}$ in the ISP factor in 
Eq.~\eqref{Eq:Coherence} fills  the role of the initial momentum and 
the four-momentum of the prepared initial state is expressed as
\begin{equation}
P^\mu=(E_{eg},\bm{p}_{eg})\,,
\end{equation}
in the rest frame of the target atomic system, and the $\delta$ function
in Eq.~\eqref{Eq:Amp} is replaced by 
$\delta^3(\bm{p}_{eg}-\bm{p}_\gamma-\bm{p}-\bm{p'})$.
The RENP process with nonvanishing $\bm{p}_{eg}$ is mentioned as boosted RENP.
We note that $\sqrt{P^2}\leq E_{eg}$, i.e. the invariant mass of 
the virtual parent particle in the boosted RENP is smaller than
that in the case of vanishing boost ($\bm{p}_{eg}=0$). 
It is expected that the boosted RENP exhibits a higher kinematic 
sensitivity to properties of neutrinos related to their masses.

The energy-momentum conservation in the boosted RENP with a trigger
laser $\gamma$ is expressed as 
\begin{equation}
q^\mu=p^\mu+p'^\mu\,, 
\end{equation}
where the four-momentum of the neutrino pair $q^\mu$ is given by 
\begin{equation}\label{Eq:qmu}
q^\mu=P^\mu-p_\gamma^\mu=(E_{eg}-E_\gamma,\bm{p}_{eg}-\bm{p}_\gamma)\,.
\end{equation}

In order for the RENP process to take place,
the invariant mass of the neutrino pair must be larger than the
sum of the masses of the emitted neutrinos:
\begin{align}
s&:=q^2=E_{eg}^2-\bm{p}_{eg}^2
        -2E_\gamma(E_{eg}-|\bm{p}_{eg}|\cos\theta_\gamma)\,,\nonumber\\
 &>(m_j+m_i)^2\,,
\end{align}
where $\theta_\gamma$ is the angle between $\bm{p}_{eg}$ and $\bm{p}_\gamma$.
The magnitude of the initial momentum $\bm{p}_{eg}$ is given by
$|\bm{p}_{eg}|=\omega_1-\omega_2$ in the coherence preparation 
scheme of the counter-propagating two-photon absorption. Here 
the convention of $\omega_1\geq\omega_2$ is employed.
We take $|\cos\theta_\gamma|=1$ assuming that the trigger photon is 
(anti)parallel to $\bm{p}_{eg}$. 
Then the trigger photon energy is expressed as
\begin{equation}
E_\gamma=\frac{1}{2}\left[E_{eg}\pm|\bm{p}_{eg}|
                          -\frac{s}{E_{eg}\mp|\bm{p}_{eg}|}\right]
=\omega_{1(2)}-\frac{s}{4\omega_{2(1)}}\,,
\end{equation}
and thus
\begin{equation}
\label{Eq:TrigE}
0<E_\gamma<\omega_{1(2)}-\frac{(m_j+m_i)^2}{4\omega_{2(1)}}\,.
\end{equation}
We note that the case of $\omega_1=\omega_2=E_{eg}/2$ is of
no boost and Eq.~\eqref{Eq:TrigE} reproduces the RENP threshold
$\omega_{ji}=E_{eg}/2-(m_j+m_i)^2/(2E_{eg})$ in 
Refs.~\cite{DinhPetcovSasaoTanakaYoshimura2012a,Fukumi2012a}.

\section{\label{Sec:Rate} Rate formula of the boosted RENP}
We present a rate formula of the boosted RENP introduced in the previous
section. 
The differential rate is written as
\begin{equation}\label{Eq:DiffRate}
 d\Gamma_{ji}=  
  n^2V \frac{(\bm{d}_{pg}\cdot\langle\rho_{eg}\bm{E}\rangle)^2}
            {(E_{pg}-E_\gamma)^2}
  \sum_{\nu\text{ hel.'s}}|\mathcal{M}_W|^2 d\Phi_2\,,
\end{equation}
where $\mathcal{M}_W$ is the weak matrix element, the two-body phase
space is given by
\begin{equation}
 d\Phi_2=(2\pi)^4\delta^4(q-p-p')\frac{d^3p}{2p^0}\frac{d^3p'}{2p'^0}\,,
\end{equation}
and $\langle\rho_{eg}\bm{E}\rangle$ represents the average of
$\rho_{eg}(\bm{x})\bm{E}(\bm{x})$ over the target with 
$\bm{E}(\bm{x})$ being the electric field in the target stimulated 
by the trigger laser. 
The single intermediate state $|p\rangle$ is assumed to dominate
with the expectation value of the dipole operator $\bm{d}_{pg}$,
and $E_{pg}:=E_p-E_g$ is introduced in the energy denominator. 
The four-momentum of the neutrino pair $p+p'$ is subject to the
four-momentum conservation dictated by the macrocoherence as shown 
by the delta function, $\delta^4(q-p-p')$. The four-momentum $q$
is given by Eq.~\eqref{Eq:qmu}.

Integrating over the neutrino phase space and summing over the mass
eigenstates, we obtain the following spectral rate,
\begin{align}
 \Gamma(E_\gamma)=&\sum_{j,i}\int d\Gamma_{ji}\nonumber\\
 =&\Gamma_0\sum_{j,i}\frac{\beta s}{6(E_{pg}-E_\gamma)^2}
   \frac{E_\gamma}{E_{eg}}
   \left[|c^A_{ji}|^2\left\{2-\frac{m_j^2+m_i^2}{s}
                             -\frac{(m_j^2-m_i^2)^2}{s^2}\right.\right.
    \nonumber\\
  &\left. +\frac{2}{3}\frac{\bm{q}^2}{s}
           \left(1+\frac{m_j^2+m_i^2}{s}
                 -2\frac{(m_j^2-m_i^2)^2}{s^2}\right)
    \right\}
    \left.-6\delta_M\mathrm{Re}(c^{A2}_{ji})\frac{m_jm_i}{s}\right]\,,
     \label{Eq:RENPrate}
\end{align}
where 
\begin{equation}
 \beta^2=1-2\frac{m_j^2+m_i^2}{s}+\frac{(m_j^2-m_i^2)^2}{s^2}\,,
\end{equation}
$c^A_{ji}:=U_{ej}^*U_{ei}-\delta_{ji}/2$ represents the neutrino mixing
factor, and $\delta_M=0(1)$ for Dirac (Majorana) neutrinos.
The overall rate $\Gamma_0$ for the target of number density $n$,
volume $V$ and dynamical activity $\eta$ is given by
\begin{equation}
\Gamma_0:=\frac{2G_F^2}{\pi}\langle\bm{s}\rangle^2 n^2V
          |\bm{d}_{pg}\cdot\langle\rho_{eg}\bm{E}\rangle|^2
          \frac{E_{eg}}{E_\gamma}
         =(2J_p+1)C_{ep}G_F^2\frac{\gamma_{pg}E_{eg}}{E_{pg}^3}n^3V\eta\,,
\end{equation}
where $\bm{s}$ is the electron spin operator,
$J_p$ is the angular momentum of the intermediate state $|p\rangle$, 
$C_{ep}$ denotes the spin matrix matrix element, 
and $\gamma_{pg}$ is the rate of the $|p\rangle\to|g\rangle$ E1 transition. 
The dynamical activity factor $\eta$ of the target is
defined by\footnote{The energy density of the trigger field
is $|\bm{E}|^2/2$ and its value is $E_\gamma n$ when each atom in
the target emits a photon of $E_\gamma$, 
while the maximal value of $|\rho_{eg}|$ is 1/2.
Hence $|\langle\rho_{eg}\bm{E}\rangle|^2\leq E_\gamma n/2$ and
$\eta\leq 1$ follows from this definition. The definition of $\eta$ 
in the present work is different from that in 
Refs.~\cite{DinhPetcovSasaoTanakaYoshimura2012a,Fukumi2012a}.}
\begin{equation}
|\langle\rho_{eg}\bm{E}\rangle|^2=:\frac{1}{2}\eta E_\gamma n\,.
\end{equation}
For both Yb~\cite{DinhPetcovSasaoTanakaYoshimura2012a} and 
Xe~\cite{Fukumi2012a}, $J_p=1$ and $C_{ep}=2/3$. Thus, we obtain
\begin{equation}
\Gamma_0=2G_F^2\frac{\gamma_{pg}E_{eg}}{E_{pg}^3} n^3V\eta\,.
\end{equation}

It is notable that we may discriminate Dirac and Majorana neutrinos
with the RENP spectrum in Eq.~\eqref{Eq:RENPrate} owing to 
the Majorana interference shown by the last term in the square brackets.
Furthermore, if the neutrinos are Majorana fermions, there appear two extra
CP violating phases in the lepton sector. We have virtually
no experimental information on these Majorana phases at present. 
One of the advantages of the boosted RENP is its good sensitivity to 
Majorana phases as we show below.

The lepton mixing matrix appearing in $c^A_{ji}$ is represented 
as a product of two unitary matrices~\cite{PDG2016}
\begin{equation}
U=VP\,,
\end{equation}
where the PMNS matrix $V$ is written in terms of three mixing angles
and the CP violating Dirac phase,
\begin{equation}
V=\left[\begin{array}{ccc}
          c_{12}c_{13} & s_{12}c_{13} & s_{13}e^{-i\delta}\\
         -s_{12}c_{23}-c_{12}s_{23}s_{13}e^{i\delta} & 
          c_{12}c_{23}-s_{12}s_{23}s_{13}e^{i\delta} &
          s_{23}c_{13}\\
          s_{12}s_{23}-c_{12}c_{23}s_{13}e^{i\delta} & 
          c_{12}c_{23}-s_{12}s_{23}s_{13}e^{i\delta} & 
          c_{23}c_{13}
        \end{array}
  \right]\,,
\end{equation}
with $c_{ij}=\cos\theta_{ij}$ and $s_{ij}=\sin\theta_{ij}$.
The diagonal unitary matrix $P$ may be expressed as
\begin{equation}
P=\mathrm{diag.}(1,e^{i\alpha},e^{i\beta})\,,
\end{equation}
for Majorana neutrinos, and we can rotate away the phases $\alpha$
and $\beta$ for Dirac neutrinos resulting in the single CP violating phase. 
In our numerical calculation, we employ 
the best-fit results of NuFIT~\cite{Esteban2017a} for the neutrino mass and
mixing parameters.

The Majorana phases affect the RENP rate in Eq.~\eqref{Eq:RENPrate} 
through the offdiagonal components of $\text{Re}(c^{A2}_{ji})$:
$\text{Re}(c^{A2}_{12})=c_{12}^2s_{12}^2c_{13}^4\cos 2\alpha$,
$\text{Re}(c^{A2}_{13})=c_{12}^2c_{13}^2s_{13}^2\cos 2(\beta-\delta)$,
$\text{Re}(c^{A2}_{23})=s_{12}^2c_{13}^2s_{13}^2\cos 2(\beta-\delta-\alpha)$.
We observe that the dependence of the RENP rate on $\beta-\delta$ 
is relatively weak because of the rather suppressed mixing angle
$s_{13}^2\sim 0.022$. The RENP experiment is complementary to 
oscillation experiments that can probe
$\delta$~\cite{AbeT2K2017a,AdamsonNOvA2017a}.

\section{\label{Sec:DM} 
         Dirac-Majorana distinction and effect of Majorana phases}

\begin{figure}
 \centering
 \includegraphics[width=0.45\textwidth]{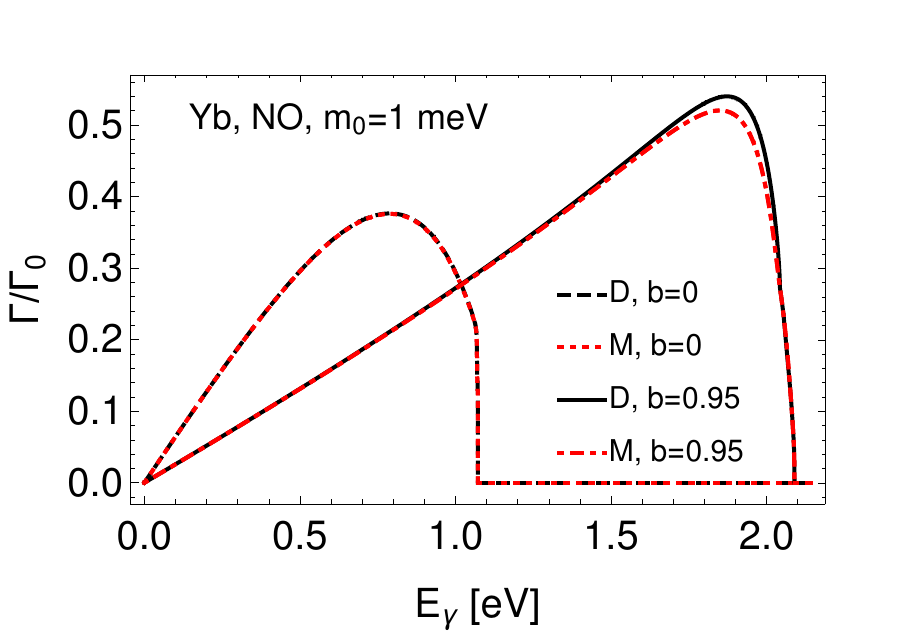}\ 
 \includegraphics[width=0.45\textwidth]{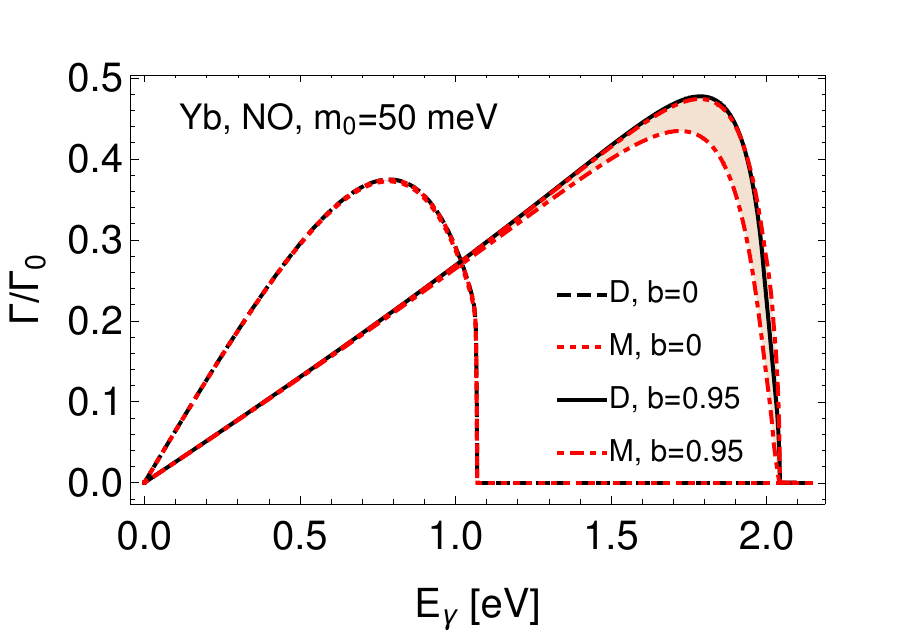}
 \caption{Dirac-Majorana difference in the spectral shape, 
          $\Gamma(E_\gamma)/\Gamma_0$, with ($b=0.95$) and without
          ($b=0$) boost.
          Yb, NO, $0<\alpha<\pi/2$ and $\beta=0$.
          The smallest neutrino mass is chosen as $m_0=$ 
          1 meV (left) and 50 meV (right).}
 \label{Fig:DMNO}
\end{figure}

\begin{figure}
 \centering
 \includegraphics[width=0.45\textwidth]{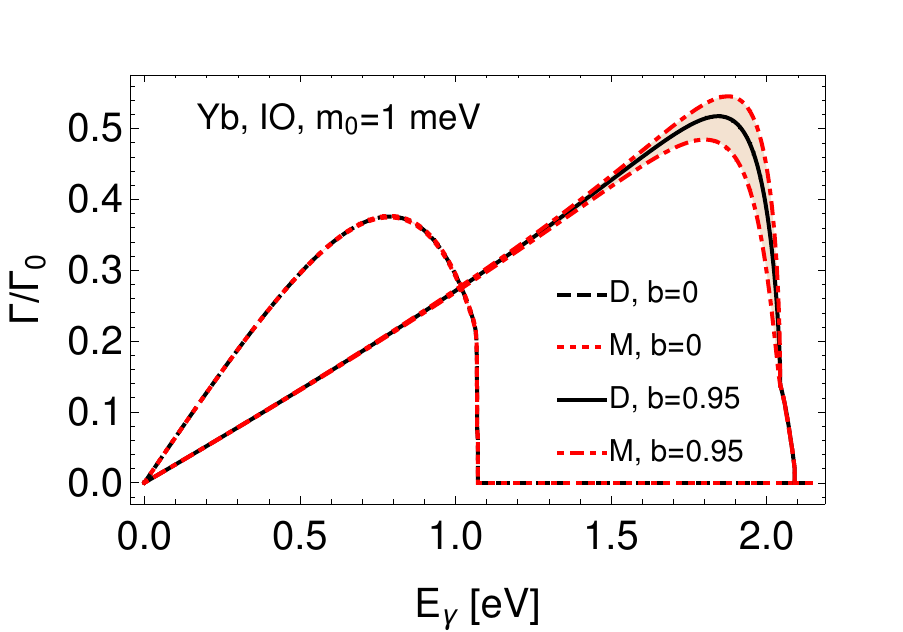}\ 
 \includegraphics[width=0.45\textwidth]{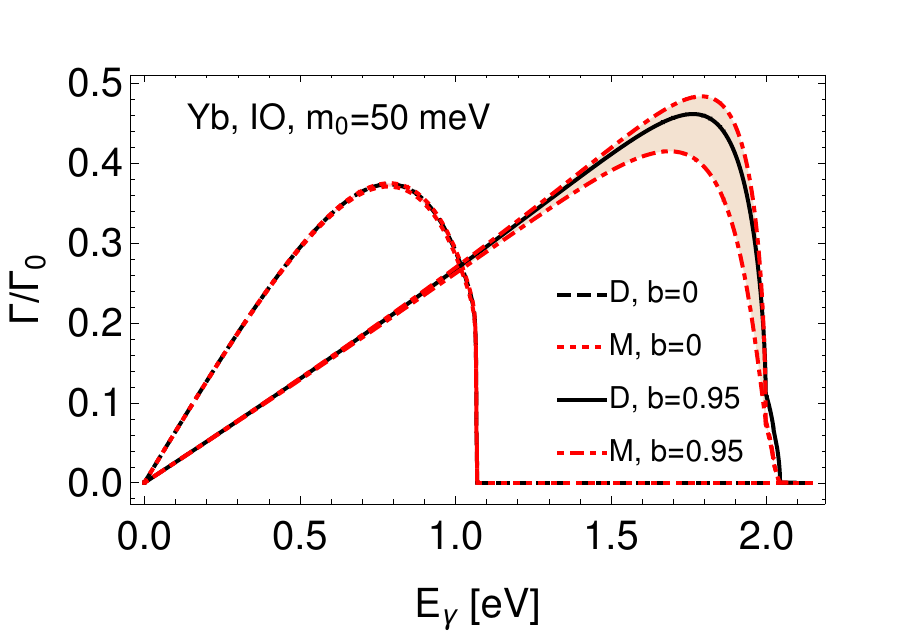}
 \caption{Dirac-Majorana difference in the IO case. The other parameters
          are the same as Fig.~\ref{Fig:DMNO}.}
 \label{Fig:DMIO}
\end{figure}

We compare the boosted RENP spectra for Dirac and Majorana neutrinos
and examine the effect of Majorana phases.
Figure \ref{Fig:DMNO} shows the spectral shape $\Gamma(E_\gamma)/\Gamma_0$
in the case of Yb (%
$|g\rangle=\text{6s}^2\,{^1\text{S}}_0$, 
$|e\rangle=\text{6s6p}\,{^3\text{P}}_0$,
$|p\rangle=\text{6s6p}\,{^3\text{P}}_1$, 
$E_{eg}=2.14348$ eV and $E_{pg}=2.23072$ eV
\cite{DinhPetcovSasaoTanakaYoshimura2012a})
for the normal ordering (NO) of neutrino masses
with the smallest neutrino mass $m_0$ being 1 meV (left) and 50 meV (right). 
The trigger is taken parallel to the ISP momentum $\bm{p}_{eg}$ and 
the boost magnitude is $b:=|\bm{p}_{eg}|/E_{eg}=0.95$.
This is realized by choosing $\omega_1=2.08989$ eV and $\omega_2=0.05359$ eV.
The black solid lines represent the spectra of the Dirac case with this boost
and the spectra without boost ($b=0$) are also shown by the black
dashed lines for comparison. The endpoint for $b=0$ is $\sim E_{eg}/2$ and
that for $b=0.95$ is close to $E_{eg}$ as given in Eq.~\eqref{Eq:TrigE}.
As for the case of Majorana neutrinos,
we vary $\alpha$ while $\beta$ is fixed to zero. The red dash-dotted lines
represent the spectra of $\alpha=0$ and $\pi/2$ and the shaded regions
corresponds to $\alpha$ between these two values. We also show the cases of
no boost as the red dotted lines for comparison. We note that 
the boundaries of $\alpha\in [0,\pi/2]$ are indistinguishable in the cases
without boost and even with the boost for $m_0=1$ meV.
The case of inverted ordering (IO) is presented in Fig.~\ref{Fig:DMIO}. 

We observe that enhancement of Dirac-Majorana difference is possible
in the boosted RENP. In particular, near the endpoint 
($E_\gamma\sim E_{eg}$), the difference becomes larger than 10 \%
although the rate itself is suppressed. 
The effect of Majorana phases is also significantly enhanced by boosting.
A sizable effect in the rate, say 10 \% or more, is expected if
$m_{1,2}\sim 50\ \text{meV}$, which is always the case in the inverted 
ordering.

\begin{figure}
 \centering
 \includegraphics[width=0.45\textwidth]{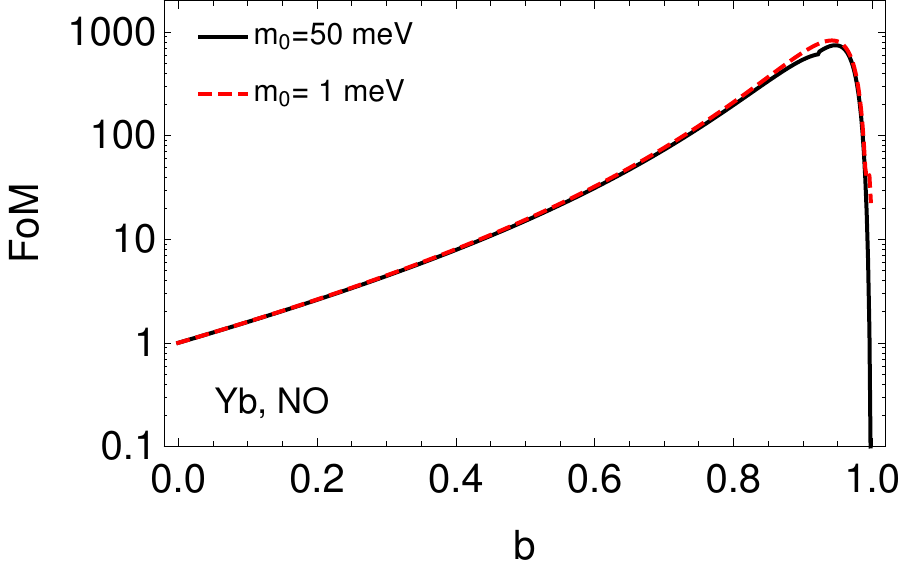}
 \hspace{0.05\textwidth} 
 \includegraphics[width=0.45\textwidth]{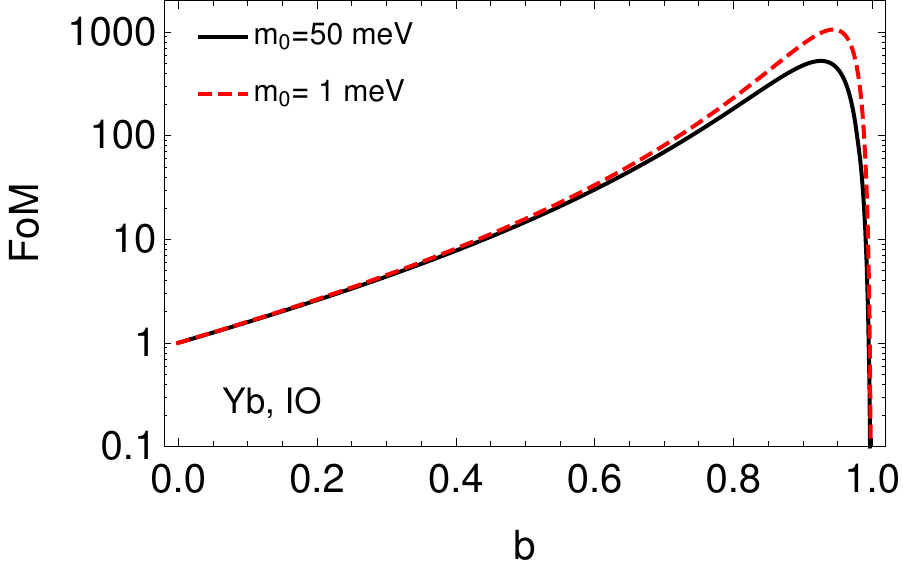}
 \caption{Maximal figure of merit as a function of the boost magnitude,
          $b=|\bm{p}_{eg}|/E_{eg}$. 
          Yb, NO (left) and IO (right), $\alpha=\beta=0$, and
          $m_0$=1 meV (red dashed) and 50 meV (black solid).}
 \label{Fig:FOMDM}
\end{figure}

In order to quantify the power of the boost by the ISP in discriminating
Dirac and Majorana cases, we introduce the following figure of
merit (FoM) function,
\begin{equation}
\mu(E_\gamma):=\frac{2A^2(E_\gamma)}{1+|A(E_\gamma)|}
               \left[\Gamma_M(E_\gamma)+\Gamma_D(E_\gamma)\right]\,,
\end{equation}
where $\Gamma_{M}(E_\gamma)$ and $\Gamma_{D}(E_\gamma)$ denote 
the Majorana and Dirac RENP rates respectively,
and the asymmetry $A(E_\gamma)$ is defined by
\begin{equation}
A(E_\gamma):=\frac{\Gamma_M(E_\gamma)-\Gamma_D(E_\gamma)}
                 {\Gamma_M(E_\gamma)+\Gamma_D(E_\gamma)}\,.
\end{equation}
To obtain the best sensitivity, it is presumed in an experiment that
the trigger energy is chosen to maximize $\mu(E_\gamma)$
for a given magnitude of the boost, $b=|\bm{p}_{eg}|/E_{eg}$. 
In Fig.~\ref{Fig:FOMDM}, we present the maximal value of $\mu(E_\gamma)$
as a function of the boost magnitude $b$ taking $\alpha=\beta=0$ for
an illustration. The ordinate is normalized so that the maximal 
figure of merit is unity for the case of no boost. 
The left and right panels show the NO and IO cases respectively.
We observe that the FoM is enhanced by almost a factor of 1000 choosing
the best boost factor.
This means that we effectively gain statistics by a factor of 
$\sim\sqrt{1000}$ using the boost.

\section{\label{Sec:CNB} Spectral distortion by cosmic neutrino background}
The standard cosmology predicts that the universe is filled with 
background neutrinos, the cosmic neutrino background (CNB).
The neutrinos in a mass eigenstate follow the distribution,
\begin{equation}
f(\bm{p})=\frac{1}{1+e^{|\bm{p}|/T-\xi}}\,,
\end{equation}
where $\bm{p}$ is the neutrino momentum, $T\simeq 1.9\ \text{K}$ 
represents the neutrino temperature, and $\xi$ denotes the neutrino
degeneracy (assumed common to the three neutrino mass eigenstates), 
whose absolute value is constrained as $O(0.1)$ or less by the primordial 
nucleosynthesis
\cite{WagonerFowlerHoyle1966a,Rana1982a,KangSteigman1991a,SchwarzStuke2012a}. 
The distribution of antineutrinos, $\bar f(\bm{p})$
is given by changing the sign of $\xi$. We take $\xi=0$ in the following
numerical calculation.

As pointed out in Ref.~\cite{YoshimuraSasaoTanaka2014a},
the RENP spectrum is distorted by the CNB owing to the Pauli principle.
The differential rate in Eq.~\eqref{Eq:DiffRate} is modified 
by the Pauli-blocking factors as 
\begin{equation}
 d\Gamma_{ji}=  
  n^2V \frac{(\bm{d}_{pg}\cdot\langle\rho_{eg}\bm{E}\rangle)^2}
            {(E_{pg}-E_\gamma)^2}
  \sum_{\nu\text{ hel.'s}}|\mathcal{M}_W|^2 
  \{1-f(\bm{p})\}\{1-\bar f(\bm{p}')\}d\Phi_2\,.
\end{equation}
The spectral rate is obtained by integrating over the neutrino phase
space and summing over the neutrino mass eigenstates,
\begin{align}
&\Gamma(E_\gamma;T,\xi)=\sum_{j,i}\int d\Gamma_{ji}\nonumber\\
&=\Gamma_0\frac{8\pi}{(E_{pg}-E_\gamma)^2}\frac{E_\gamma}{E_{eg}}
  \sum_{j,i}\int d\Phi_2\{1-f(\bm{p})\}\{1-\bar f(\bm{p}')\}\times\nonumber\\
&\phantom{\Gamma_0}\left[|c^A_{ji}|^2\left\{\frac{2}{3}\bm{p}\cdot\bm{p}'+
                                            \frac{1}{2}(s-m_j^2-m_i^2)\right\}
                         -\delta_M\mathrm{Re}(c^{A2}_{ji})m_j m_i\right]\,.
\end{align}

\begin{figure}
 \centering
 \includegraphics[width=0.45\textwidth]{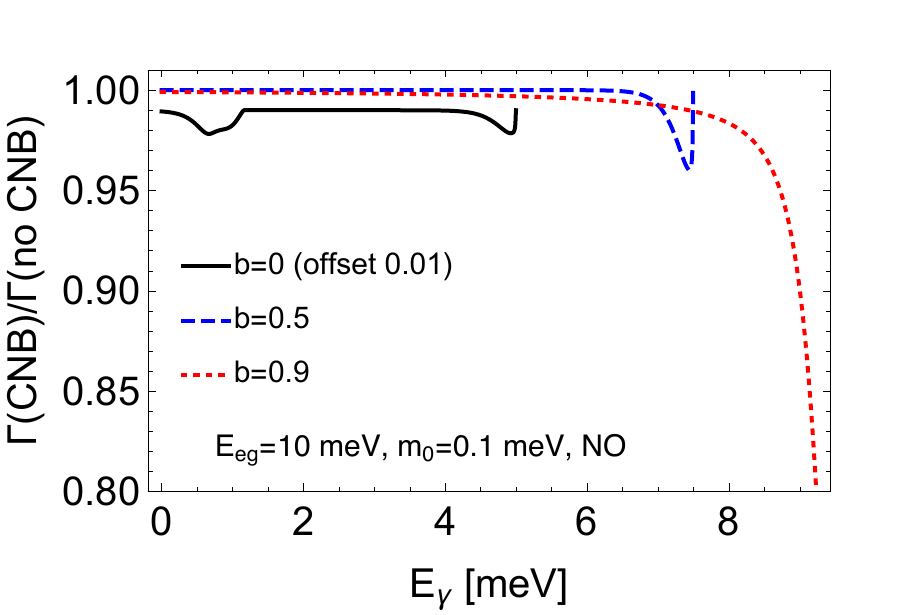}
 \hspace{0.05\textwidth} 
 \includegraphics[width=0.45\textwidth]{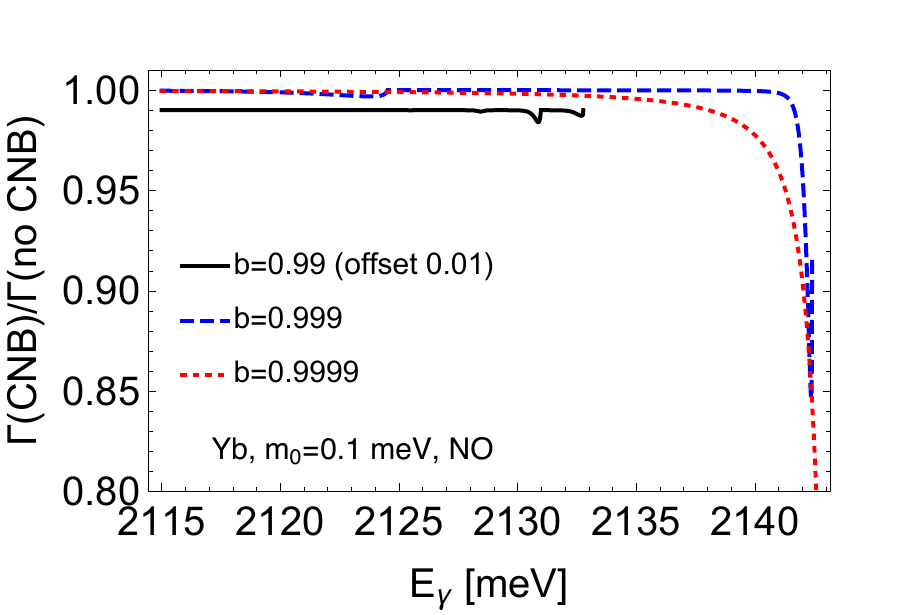}
 \caption{RENP spectral distortion by the CNB. 
          The ordinate is the ratio of RENP rates with and without
          the Pauli blocking by the CNB.
          The neutrino parameters are 
          $m_0=0.1$ meV, NO, Majorana and $\alpha=\beta=0$. 
          Left: $E_{eg}=10$ meV, $E_{pg}=1$ eV, no boost (black solid), 
                50\% boost (blue dashed) and 90\% boost (red dotted).
          Right: Yb ($E_{eg}=2.14348$ eV, $E_{pg}=2.23072$ eV), 
                 99\% boost (black solid), 
                 99.9\% boost (blue dashed) and 99.99\% boost (red dotted).
                 The black lines are offset by 0.01 for better separations.}
 \label{Fig:CNB}
\end{figure}

We illustrate possible spectral distortions in Fig.~\ref{Fig:CNB}
for the cases of a hypothetical atom with a very small level
splitting, $E_{eg}=10$ meV (left) and Yb (right).
The ordinate is the ratio of RENP rates with and without the Pauli blocking
by the CNB.
The mass of the lightest neutrino is chosen to be $0.1$ meV.
Although the spectral distortion is sizable in the case of no boost
for the tiny level splitting, an appropriate boost substantially
enhances the distortion for the both cases in Fig.~\ref{Fig:CNB}.
Effects of 10\% or more are expected near the endpoints.

\section{\label{Sec:Conclusion} Conclusion}
We have explored the effect of the initial spatial phase (ISP)
in the radiative emission of neutrino pairs (RENP). 
The ISP is provided by the two-photon absorption with two lasers of 
different frequencies in the preparation process of the coherent initial
state of a macroscopic target. The ISP factor is interpreted to give 
a momentum $\bm{p}_{eg}$ to the initial state of RENP, 
so that the RENP process with the ISP is called as boosted RENP.
Owing to the momentum conservation dictated by the macrocoherent rate 
enhancement mechanism, $\bm{p}_{eg}$ changes the kinematics of the RENP
process as if the invariant mass of the parent particle decreases. 
This effective reduction of the energy scale makes the RENP process 
kinetically more sensitive to the emitted neutrino masses.

We have evaluated the effect of the ISP in the RENP spectra.
It is shown that the difference between the Dirac and Majorana neutrinos 
is significantly enhanced in the boosted RENP as presented in 
Figs.~\ref{Fig:DMNO} and \ref{Fig:DMIO}.
The figure of merit function in Fig.~\ref{Fig:FOMDM} shows that the best
choice of the boost factor provides us a statistical merit of $O(10)$.
In addition, the possible spectral distortion by the cosmic neutrino
background is investigated. As shown in Fig.~\ref{Fig:CNB}, 
the spectral distortion becomes more substantial in the boosted RENP.

For improved capability of the Dirac-Majorana distinction and
the CNB detection, it is vital
to incorporate the ISP effect (or the boost) in the design 
of the RENP experiment. 
The SPAN collaboration has already observed the signal of the paired 
superradiance (PSR) from a parahydrogen target in the two-photon 
absorption scheme~\cite{SPAN201X}.
They use two identical counter-propagating lasers at present, 
so that no ISP is generated. 
After establishing the preparation of the initial coherent state 
by the two-photon absorption, the PSR with an ISP (boosted PSR)
becomes possible as a prototype of the boosted RENP.

\section*{Acknowledgments}
This work is supported in part by JSPS KAKENHI Grant Numbers 
JP 15H02093, 15H03660, 15K13468, 16H00868, 16H03993, 17H02895 and 17H05405.

\end{document}